\documentclass[graybox]{svmult}

\smartqed
\usepackage{mathptmx}       
\usepackage{helvet}         
\usepackage{courier}        
\usepackage{type1cm}        
\usepackage{graphicx}       

\usepackage{array,colortbl}
\usepackage{amsmath,amsfonts,amssymb,bm} 
\usepackage{multirow}
\usepackage{booktabs} 
\usepackage{pdflscape}
\newcolumntype{H}{>{\setbox0=\hbox\bgroup}c<{\egroup}@{}}

\usepackage{listings}
\DeclareSymbolFont{bbold}{U}{bbold}{m}{n}
\DeclareSymbolFontAlphabet{\mathbbold}{bbold}

\usepackage{microtype} 

\usepackage[colorlinks=true,linkcolor=black,citecolor=black,urlcolor=black]{hyperref}
\usepackage{bookmark}
\makeatletter 
  \providecommand*{\toclevel@author}{999}
  \providecommand*{\toclevel@title}{0}
\makeatother

\usepackage{tikz}
\usetikzlibrary{shapes,decorations,arrows,automata,patterns,plotmarks,petri}
\usepackage{pgfplots}
\pgfplotsset{compat=1.12}
\usepackage{tikz-inet}
\usepackage{xcolor}
\usepackage{caption}
\usepackage{subcaption}
\usepackage{etoolbox}

{\obeyspaces\gdef {\ }}
\def\setverbatim{\def\par{\leavevmode\endgraf}
            \parskip=0pt\parindent=0pt\obeylines\obeyspaces }
\chardef\other=12
\def\ttverbatim{\setverbatim\tt
       \catcode`\{=\other \catcode`\}=\other \catcode`\_=\other
       \catcode`\^=\other \catcode`\$=\other \catcode`\%=\other
       \catcode`\#=\other \catcode`\&=\other \baselineskip=11pt
       }

\def\code {\vfil\vfilneg\vbox\bgroup\ttverbatim}
\def\longcode {\vfil\vfilneg\bgroup\ttverbatim}

\def\footc {\footnotesize\tt\baselineskip=9.0pt}

\def\smallcode {\code\footc\renewcommand\ttfamily{\fontfamily{ulg}\selectfont}}

\let\endcode=\egroup

\def\bm#1{\mathbf{#1}}

\def\cD   {\mathcal{D}}

\def\cH   {\mathcal{H}}

\def\cO   {\mathcal{O}}
\def\cP   {\mathcal{P}}

\def\cR   {\mathcal{R}}

\def\cV   {\mathcal{V}}

\def\EE{\mathbb{E}}

\def\II{\mathbb{I}}

\def\RR{\mathbb{R}}
\def\NN{\mathbb{N}}

\def\ZZ{\mathbb{Z}}
\def\LL{\mathbb{L}}
\def\?{\discretionary{}{}{}}  

\newcommand{\fraku}{{\mathfrak{u}}}

\newcommand{\Var}{\mathrm{Var}}

\definecolor{vector}{cmyk}{0.0,0.8,1.0,1.0}  
\definecolor{tan}{cmyk}{0.30,0.50,0.60,0}
\definecolor{orange}{cmyk}{0.0,0.6,1.0,0.1}
\definecolor{names}{cmyk}{1.0,0.0,1.0,0.14}
\definecolor{pink}{cmyk}{0.0,0.8,0,0}
\definecolor{paleyellow}{cmyk}{0,0,0.6,0.0}
\definecolor{darkyellow}{cmyk}{0,0.2,1.0,0.2}

\usepackage[utf8]{inputenc}
\usepackage{xcolor}

\begin{document}

\title*{A Tool for Custom Construction of QMC and RQMC Point Sets}

\author{Pierre L'Ecuyer \and Pierre Marion \and Maxime Godin \and Florian Puchhammer}

\institute{
 Pierre L'Ecuyer
 \at D\'epartement d'Informatique et de Recherche Op\'erationnelle,
     Universit\'e de Montr\'eal, Canada,
 \email{lecuyer@iro.umontreal.ca}
\and
Pierre Marion
 \at Sorbonne Universit\'e, CNRS, Laboratoire de Probabilités, Statistique et Modélisation, LPSM, \mbox{F-75005} Paris, France, \email{pierre.marion@upmc.fr}
\and
 Maxime Godin
 \at Institut des Actuaires, France, \email{maxime.godin@institutdesactuaires.com}
\and
 Florian Puchhammer
 \at Universit\'e de Montr\'eal, Canada, and Basque Center for Applied Mathematics, Spain,
     \email{fpuchhammer@bcamath.org}
}

\maketitle

\abstract*{
We present LatNet Builder, a software tool to find good parameters for lattice rules, 
polynomial lattice rules, and digital nets in base 2, for quasi-Monte Carlo (QMC) and
randomized quasi-Monte Carlo (RQMC) sampling over the $s$-dimensional unit hypercube.
The selection criteria are figures of merit that give different weights to different subsets of 
coordinates.
They are upper bounds on the worst-case error (for QMC) or variance (for RQMC)
for integrands rescaled to have a norm of at most one in certain Hilbert spaces of functions
We summarize what are the various Hilbert spaces, discrepancies, types of weights,
figures of merit, types of constructions, and search methods
supported by LatNet Builder.  We briefly discuss its organization and 
we provide simple illustrations of what it can do.
}

\abstract{
We present LatNet Builder, a software tool to find good parameters for lattice rules, 
polynomial lattice rules, and digital nets in base 2, for quasi-Monte Carlo (QMC) and
randomized quasi-Monte Carlo (RQMC) sampling over the $s$-dimensional unit hypercube.
The selection criteria are figures of merit that give different weights to different subsets of 
coordinates.
They are upper bounds on the worst-case error (for QMC) or variance (for RQMC)
for integrands rescaled to have a norm of at most one in certain Hilbert spaces of functions.
We summarize what are the various Hilbert spaces, discrepancies, types of weights,
figures of merit, types of constructions, and search methods
supported by LatNet Builder.  We briefly discuss its organization and  
we provide simple illustrations of what it can do.
}

\section{Introduction}
\label{sec:intro}

QMC methods approximate an integral of the form
\begin{equation}
 \mu = \int_0^1 \cdots \int_0^1 f(u_1,\dots,u_s)\, du_1 \cdots du_s
     = \int_{(0,1)^s} f({\bm u})\, d{\bm u}
     = \EE[f({\bm U})]                   \label{eq:intf}
\end{equation}
where $f: (0,1)^s\to \RR$ and $\bm U$ is a uniform random vector over $[0,1)^s$, by the average 
\begin{equation}
  \bar\mu_n = \frac{1}{n} \sum_{i=0}^{n-1} f({\bm u}_i)   \label{eq:aver-qmc}
\end{equation}
where $P_n = \{{\bm u}_0,\dots,{\bm u}_{n-1}\} \subset [0,1)^s$ is a set
of $n$ deterministic points that cover the unit hypercube more evenly than 
typical independent random points.
That is, the discrepancy between their empirical distribution and the uniform distribution
over $[0,1)^s$ is smaller than for independent random points and converges to 0 faster
than $\cO(n^{-1/2})$ when $n\to\infty$.
This discrepancy can be defined in many ways.  It usually represents the worst-case 
integration error for a given class of integrands $f$.
Typically, this class is a reproducing-kernel Hilbert space (RKHS) ${\cH}$ of functions, such that
\begin{equation}
  |E_n| := |\bar\mu_n - \mu| \le \cD(P_n) \cV(f)       \label{eq:HK}
\end{equation}
for all $f\in\cH$, where $\cV(f)$ is the norm of $f-\mu$ in $\cH$
(we call it the \emph{variation} of $f$)
and $\cD(\cdot)$ is the \emph{discrepancy} measure associated with this Hilbert space
\cite{rDIC10a,vHIC98a,rNIE92b}.
For a fixed $f\in{\cal H}$ with $\cV(f) > 0$, the error bound in (\ref{eq:HK}) 
converges at the same rate as $\cD(P_n)$.
A traditional version of (\ref{eq:HK}), whose derivation does not involve Hilbert 
spaces, is the classical Koksma-Hlawka inequality,
in which $\cV(f)$ is the Hardy-Krause variation and $\cD(P_n)$ is the star discrepancy, 
which converges as $\cO((\log n)^{s-1} n^{-1})$ for well-selected point sets \cite{rNIE92b}.
Another important choice for $\cH$ is a Sobolev space of functions whose mixed partial derivatives
of order up to $\alpha$ are square-integrable.
It is known that for this space, one can construct point sets
whose discrepancy converges as $\cO((\log n)^{(s-1)/2} n^{-\alpha})$,
and that this is the best possible rate \cite{vDIC08a,rDIC10a,rGOD19a,rGOD16a,rGOD17a,rGOD18a}.
The main classes of QMC point sets are lattice points and digital nets.

For RQMC, the $n$ QMC points are randomized to provide a set of random points
$\{{\bm U}_0,\dots,{\bm U}_{n-1}\} \subset (0,1)^s$ for which
(i) each $\bm U_i$ individually has the uniform distribution over $[0,1)^s$, and
(ii) the points keep their highly-uniform distribution collectively.
Randomizations that provably preserve the low discrepancy generally depend on the type 
of QMC construction: some are used for lattice points and others for digital nets. 
In some cases, the randomization may even improve the convergence rate of the 
mean square discrepancy.  The RQMC estimator
\begin{equation}                        \label{eq:rqmc}
 {\hat\mu_{n,{\rm rqmc}}} = \frac{1}{n} \sum_{i=0}^{n-1} f({\bm U}_i),
\end{equation}
which is now random, is unbiased for $\mu$ and one wishes to minimize its variance.
For more details on RQMC, see for example 
\cite{vHIC01d,vLEC09f,vLEC00b,vLEC02a,rMAT98c,vOWE97a,vOWE97b,vOWE98a}.

The aim of this paper is to introduce \emph{LatNet Builder},
a software tool designed to construct good lattice and digital point sets for QMC and RQMC,
in any number of dimensions, for an arbitrary number of points, 
arbitrary weights on the subsets of coordinates, arbitrary smoothness of the integrands,
a variety of construction and randomization methods, and several choices of discrepancies.
The point sets can also be extensible in the number of points and number of dimensions.
By ``{constructing the points}'' here we mean \emph{defining} the set $P_n$ by selecting
the parameter values for a general structure, by trying to miminize a \emph{figure of merit} (FOM)
that may represent a discrepancy $\cD(P_n)$ or be an upper bound on it.
Once this is done, other software can be used to randomize and generate the points
for their utilization in applications; see \cite{iLEC16j,iNUY20m} for example. 
LatNet Builder is available in open source at 
\url{https://github.com/umontreal-simul/latnetbuilder}.
It is a descendant of \emph{Lattice Builder} \cite{vLEC16a},
whose scope was limited to ordinary lattice rules.
Another related tool is Nuyens' fast CBC constructions \cite{iNUY12a}.

The rest of this paper is organized as follows.
In Section~\ref{sec:point-sets}, we recall the types of QMC point sets covered by
our software, namely ordinary lattice points, polynomial lattice points,
digital nets, and their higher-order and interlaced versions, as well as the 
main randomization methods to turn these point sets into RQMC points.
The discrepancies that we consider often provide upper bounds on the mean square 
integration error when using these randomizations, for certain classes of functions.
In Section~\ref{sec:weights}, we give the general form of weighted RKHS
used in this paper and the corresponding generalized Koksma-Hlawka inequality.
We also recall the common types of weights, all supported by the software.
In Section~\ref{sec:fom}, we review and justify the various discrepancies that are 
supported by LatNet Builder and can be used as FOMs
to select the parameters of point set constructions.
In Section~\ref{sec:search-methods}, we summarize the search methods implemented in our software.
In Section~\ref{sec:illustrations}, we compare FOM values obtained by various
point set constructions and search methods. We also compare RQMC variance for simple integrands $f$.
Section~\ref{sec:conclusion} gives a conclusion.

\section{Point Set Constructions and Randomizations}
\label{sec:point-sets}

LatNet Builder handles ordinary rank-1 lattice points as well as digital nets,
which include polynomial lattice rules and high-order and interlaced constructions.

For a \emph{rank-1 lattice rule}, the point set is
\[
 P_n = \left\{{\bm u}_i = i {\bm v}_1 \bmod 1, \, i=0,\dots,n-1\right\}
\]
where $n {\bm v}_1 = {\bm a} = (a_1,\dots,a_s) \in \ZZ_n^s  \equiv \{0,\cdots, n-1\}^s$.
It is a \emph{Korobov rule} if ${\bm a} = (1, a, a^2 \bmod n, \dots, a^{s-1} \bmod n)$ 
for some integer $a \in\ZZ_n$.
The parameter to select here is the vector $\bm a$, for any given $n$.
The usual way to turn a lattice rule into an RQMC point set is by a random shift:
generate a single random point ${\bm U}$ uniformly in $(0,1)^s$, and add it to each 
point of $P_n$, modulo 1, coordinate-wise. This satisfies the RQMC conditions.
For more details on lattice rules and their randomly-shifted versions, see 
\cite{vHIC98c,vHIC02a,vLEC09f,vLEC00b,vLEC12a,vSLO94a}.

The \emph{Digital nets in base 2} handled by LatNet Builder are defined as follows. 
The number of points is $n = 2^k$ for some integer $k$.
We select an integer $w\ge k$ and $s$ generating matrices ${\bm C}_1,\dots, {\bm C}_s$ 
of dimensions $w\times k$ and of rank $k$, with elements in $\ZZ_2 \equiv \{0,1\}$.  
The points ${\bm u}_i$, $i=0,\dots,2^k-1$, are defined as follows:
for $i = a_{i,0} + a_{i,1} 2 + \cdots + a_{i,k-1} 2^{k-1}$, we take
\begin{eqnarray*}
  \begin{pmatrix} u_{i,j,1}\cr \vdots \cr u_{i,j,w}\cr \end{pmatrix} 
    &=& {\bm C}_j
        \begin{pmatrix} a_{i,0}\cr \vdots \cr a_{i,k-1}\cr \end{pmatrix}
        \bmod 2, \qquad
  u_{i,j} = \sum_{\ell=1}^w u_{i,j,\ell} 2^{-\ell}, \qquad
\end{eqnarray*}
and ${\bm u}_i = (u_{i,1},\dots,u_{i,s})$.
There are more general definitions in \cite{rDIC10a,rNIE92b}.
The parameters to optimize are the elements of the matrices ${\bm C}_j$.
Since each ${\bm C}_j$ has rank $k$, each one-dimensional projection truncated to its 
first $k$ digits is $\ZZ_n/n = \{0, 1/n, \dots, (n-1)/n\}$. 
The ordinary digital nets constructed by LatNet Builder often have $w=k$,
so the points have only $k$ digits, but this is not always true.

The most popular digital net constructions are still the \emph{Sobol' points} \cite{rSOB67a}, 
in base $b=2$, with $k\times k$ generating matrices that are upper triangular and invertible.
These matrices are constructed by a specific method, but the bits of the first few columns
above the diagonal can be selected arbitrarily, and their choice has an impact on the 
quality of the net. General-purpose choices have been proposed in \cite{rJOE08a,iLEM04a},
e.g., based on the uniformity of two-dimensional projections.
LatNet Builder allows one to construct the matrices based on a much more flexible class of criteria.

A \emph{polynomial lattice rule} (PLR) in base 2 with $n = 2^k$ points is defined as follows.
We denote by $\ZZ_2[z]$ the ring of polynomials with coefficients in $\ZZ_2$,
by $\LL_2$ the set of formal series of the form $\sum_{\ell = \ell_0}^{\infty} x_\ell z^{-\ell}$ 
with each $x_\ell\in\ZZ_2$ and $\ell_0 \in \ZZ$, and for any given integer $w \ge k$,
we define $\varphi_w : \LL_2\to\RR$ by
\begin{equation}
  \varphi_w\left(\sum_{\ell = \ell_0}^{\infty} x_\ell z^{-\ell}\right) 
	 = \sum_{\ell = \max(\ell_0, 1)}^{w} x_\ell 2^{-\ell}.
\end{equation}
We select a \emph{polynomial modulus} $Q = Q(z) \in \ZZ_2[z]$ of degree $k$, 
and a \emph{generating vector} $\bm a(z) = (a_1(z),\dots,a_s(z))\in \ZZ_2[z]^s$,
whose coordinates are polynomials of degrees less than $k$ having no common factor
with $Q(z)$. The point set of cardinality $n = 2^k$ is
{
\begin{equation}
\label{eq:polynomial-lattice-rule}
  P_n = \left\{\left(\varphi_w\left(\frac{h(z) a_1(z)}{Q(z)}\right), \dots, 
	            \varphi_w\left(\frac{h(z) a_s(z)}{Q(z)}\right)\right) 
				: h(z)\in\ZZ_2[z] \mbox{, deg}(h(z)) < k\right\}.
\end{equation}}%
Here, we want to optimize the vector $\bm a(z)$.
This point set turns out to be a digital net in base 2 whose generating matrices ${\bm C}_j$
contain the first $w$ digits of the binary expansion of the $a_j(z)/ Q(z)$.
These are Hankel matrices: each row is the previous one shifted to the left by one position,
with the last entry determined by the recurrence with characteristic polynomial $Q(z)$,
applied to the entries of the previous row.
In theory, they have an infinite number of rows, but in practice we truncate them 
to $w \ge k$ rows.  This finite $w$ should be as large as possible to obtain a good
approximation of the true PLR points. Typically, $w=31$, but it could be $w=63$ if we use 64-bit integers.
See \cite{rDIC10a,vLEC04a,vLEM03a,rNIE92b,rNIE92c} for further details on PLRs.

A \emph{high-order polynomial lattice rule} (HOPLR) of order $\alpha$ with $n = 2^k$ points 
is obtained by constructing an ordinary PLR with polynomial modulus $\tilde{Q}(z)$ of degree 
$\alpha k$ having $2^{\alpha k}$ points in $s$ dimensions, 
and using only the first $n = 2^k$ points. 
See \cite{rBAL11b,rBAL12a,rDIC07a}.
This type of construction can achieve a higher order of convergence for the error
(almost $\cO(n^{-\alpha})$)
than an ordinary PLR for integrands $f$ in a Sobolev space of smoothness order $\alpha$ 
(i.e., when all mixed partial derivatives of up to order $\alpha$ are square integrable).
One drawback is that because of the high degree
of $\tilde{Q}$, the cost of a full CBC construction (see Section~5)
is much higher since there are $2^{\alpha k}$
possibilities to examine each time we select a new coordinate of the generating vector.

Dick \cite{mDIC07a,vDIC08a} also proposed an \emph{interlacing construction}, 
for digital nets in general (which includes PLRs),
that can provide the higher-order convergence rate of almost $\cO(n^{-\alpha})$ 
for the integration error, for integrands with smoothness order $\alpha$.
For an interlacing factor $d \in\NN$, the method first constructs a digital net in 
$sd$ dimensions, with generating matrices $C_1,\dots,C_{sd}$.
Then the generating matrices of the $s$-dimensional interlaced net are $C_1^{(d)},\dots, C_s^{(d)}$,
where the rows of $C_j^{(d)}$ are the first rows of $C_{(j-1)d+1},\dots, C_{jd}$ in this order,
then the second rows of these matrices in the same order, and so on.

The simplest way to define a RQMC point set from a digital net in base 2 is to add a digital
random shift modulo 2 to all the points.  
To do this, we generate a single point ${\bm U} = (U_1,\dots,U_s)$ uniformly in $(0,1)^s$, 
and perform a bitwise exclusive-or (XOR) between the binary digits of ${\bm U}$ 
and the corresponding digits of each point ${\bm u}_i$. 

A more involved randomization method for digital nets is the 
\emph{nested uniform scramble} (NUS) of Owen \cite{vOWE97a,vOWE97b}.
In base 2, for each coordinate, we do the following.
With probability 1/2, flip the first bit of all the points.
Then, for the points whose first bit is 1, with probability 1/2, flip all the second bits.
Do the same for the points whose first bit is 0, independently.
Then do this recursively for all the bits.
After all flipping is done for the first $\ell$ bits, partition the points in $2^{\ell}$ 
batches according to the values of their first $\ell$ bits, and for each batch,
flip bit $\ell+1$ of all the points with probability 1/2, independently across the batches.
This requires $(2^{\ell}-1)s$ random bits to flip the first $\ell$ bits of all coordinates.
One can equivalently do this only for the first $k$ bits, and generate the other bits
randomly and independently across points \cite{rMAT98c}.

A less expensive scramble, which gives less independence than NUS but more than 
a digital random shift, is a (left) \emph{linear matrix scramble} (LMS) followed
by a digital random shift (LMS+shift) \cite{vHIC01d,vHON03a,rMAT98c,vOWE03a}. 
The LMS replaces ${\bm C}_j$ by $\tilde{\bm C}_j = {\bm L}_j {\bm C}_j \? \bmod 2$, where ${\bm L}_j$
is a random non-singular lower-triangular $w\times w$ binary matrix.

Owen \cite{vOWE97b} proved that under sufficient smoothness conditions on $f$,
the RQMC variance with NUS on digital nets with fixed $s$ and bounded $t$
converges as ${\cal O}(n^{-3}(\log n)^{s-1})$.
A variance bound of the same order was shown for LMS+shift in \cite{vHIC01d,vYUE02a}.
Note that these results were proved under the assumption that $w=\infty$.

\section{Hilbert Spaces and Projection-Dependent Weights}
\label{sec:weights}

The FOMs used by LatNet Builder are based on generalized (weighted) 
Koksma-Hlawka inequalities of the form (\ref{eq:HK})
where
\begin{equation}
\label{eq:weightedVariation}
  \cV^p(f) = \sum_{\emptyset\not=\fraku\subseteq\{1,2,\dots,s\}}
	           \gamma_{\fraku}^{-p} \cV^p(f_{\fraku}) \
\end{equation}
and 
\begin{equation}
\label{eq:weightedFOM}
	\cD^q(P_n) = \sum_{\emptyset\not=\fraku\subseteq\{1,2,\dots,s\}}
	             \gamma_{\fraku}^{q} \cD^q_{\fraku}(P_n),
\end{equation}
where $1/p + 1/q = 1$, 
$\gamma_{\fraku} \in\RR$ is a weight assigned to the subset $\fraku$, 
$\cV(f_{\fraku})$ is the variation of $f_{\fraku}$, 
$\cD_{\fraku}(P_n)$ is the discrepancy of the projection of $P_n$ 
over the subset $\fraku$ of coordinates,
and $f = \sum_{\fraku\subseteq\{1,2,\dots,s\}} f_{\fraku}$ is the 
functional ANOVA decomposition of $f$ \cite{tEFR81a,vOWE98a}.
LatNet Builder allows any $q\in [1,\infty]$. 
Taking $q=\infty$ with $p=1$ means removing the $q$ and taking the max instead of the sum 
in (\ref{eq:weightedFOM}), while $p=\infty$ with $q=1$ means removing the $p$ and taking 
the max instead of the sum in (\ref{eq:weightedVariation}).
The most common choice is $p=q=2$.

LatNet Builder implements 
a variety of choices for $\cD_{\fraku}(P_n)$, depending on the point set constructions.
Some of these measures correspond to the worst-case error in some function space,
assuming that the points of $P_n$ are not randomized.
Others correspond to the mean-square error (or variance), assuming that the points are 
randomized in some particular way.
This is typically done by defining a RKHS with a kernel that is invariant with respect 
to the given randomization (i.e., digital shift-invariant, scramble-invariant, etc.),
and taking the worst-case error in that space.  

The role of the weights is to better recognize the importance of the subsets $\fraku$ 
for which $f_{\fraku}$ contributes the most to the error or variance.
That is, if $\cV(f_{\fraku})$ is unusually large, we want to divide it by a larger weight
$\gamma_{\fraku}$ to control its contribution to $\cV(f)$, but then we have to multiply
$\cD_{\fraku}(P_n)$ in (\ref{eq:weightedFOM}) by the same weight.
The final effect is that the FOM will penalize more the discrepancy 
for that particular projection. 

In principle, the weights $\gamma_{\fraku}$ can be arbitrary. 
But for large $s$, defining arbitrary individual weights for the $2^s-1$ projections 
is impractical, so special forms of weights that are parameterized by much fewer than $2^s-1$ 
parameters have been proposed.
The most common ones are \emph{product weights}, for which a weight $\gamma_j$ is assigned to 
coordinate $j$ for $j=1,\dots,s$, and $\gamma_{\fraku} = \prod_{j\in\fraku} \gamma_j$;
\emph{order-dependent weights}, for which $\gamma_{\fraku} = \Gamma_{|\fraku|}$
where $\Gamma_1,\dots,\Gamma_s$ are selected constants and $|\fraku|$ is the cardinality of of $\fraku$;
and the \emph{product-and-order-dependent (POD) weights},
which are a combination of the two, with 
$\gamma_{\fraku} = \Gamma_{|\fraku|} \prod_{j\in\fraku} \gamma_j$.
These are all available in LatNet Builder.
For more discussion on how to select the weights, see
\cite{rDIC10a,vGIL18a,vLEC12b,vLEC12a,vLEC16a}, for example.

LatNet Builder can construct point sets that are extensible in the number of dimensions
and also in the number of points, which means that we can construct point sets that perform
well in the first $s$ dimensions for $s=s_{\min},\dots,s_{\max}$, and/or if we take the first
$n$ points for $n = n_1,n_2,\dots,n_m$, simultaneously.  
Typically, one would take $n_j = 2^{k_{\min} + j -1}$ for $j=1,\dots,m$,
so $n_m = 2^{k_{\max}} = 2^{k_{\min}+m-1}$ \cite{vHIC01a}.
The global FOM in this case will be a weighted sum or maximum of the FOMs over the considered 
dimensions $s$ and/or cardinalities $n_j$. 
The CBC construction approach described in Section~\ref{sec:search-methods} already 
gives a way to implement the extension in $s$.
For the extension in $n$ (or $k$), LatNet Builder implements criteria and heuristic
search methods that account for a global FOM.

\section{Figures of Merit}
\label{sec:fom}

In this section, we review the FOMs implemented in LatNet Builder.
Most of them have the general form
(\ref{eq:weightedFOM}) where typically, when the points have the appropriate special 
structure of a lattice, polynomial lattice, or digital net, and with an adapted FOM, we have
\begin{equation}
\label{eq:FOMfraku}
	\cD^q_{\fraku}(P_n) = \frac{1}{n} \sum_{i=0}^{n-1} \prod_{j\in\fraku} \phi(u_{i,j})
\end{equation}
for some function $\phi : [0,1)\to \RR$.
With product weights $\gamma_{\fraku} = \prod_{j\in\fraku} \gamma_j$, this becomes
\[
  \cD^q(P_n) = -1 + \frac{1}{n} \sum_{i=0}^{n-1} \prod_{j=1}^s (1 + \gamma_j^q \phi(u_{i,j})),
\]
which can be computed with $\cO(ns)$ evaluations of $\phi$.

As an illustration, for a randomly-shifted lattice rule, the variance is:
\begin{equation}                                 \label{eq:lattice-variance}
 \Var[\hat\mu_{n,{\rm rqmc}}] =  \sum_{{\bm 0}\not={\bm h}\in L_s^*} |\hat f({\bm h})|^2,
\end{equation}
where $L_s^* \subset \ZZ^s$ is the \emph{dual lattice} \cite{vLEC00b}.
It is also known that for periodic continuous functions having 
square-integrable mixed partial derivatives up to order ${\alpha/2}$
for an even integer $\alpha \ge 2$, one has
$|\hat f({\bm h})|^2 = {\cal O}((\max(1,h_1) \cdots \max(1,h_s))^{-\alpha})$.
This motivates the well-known FOM \cite{vLEC12a,rNIE92b,vSLO94a}:
\begin{eqnarray}
  {\cal P}_{\alpha}
   &:=& \sum_{{\bm 0}\not={\bm h}\in L_s^*} (\max(1,h_1) \cdots \max(1,h_s))^{-\alpha} \nonumber\\
   &=&  \frac{1}{n} \sum_{i=0}^{n-1} \sum_{\emptyset\not=\mathfrak{u}\subseteq\{1,\dots,s\}} 
        \left(\frac{-(-4\pi^2)^{\alpha/2}}{\alpha!}\right)^{|\fraku|} 
				  \prod_{j\in\fraku} B_{\alpha}(u_{i,j})  \label{eq:Palpha}  
\end{eqnarray}
where $B_{\alpha/2}$ is the Bernoulli polynomial of degree $\alpha/2$
($B_1(u)=u-1/2$, $B_2(u) = u^2-u+1/6$, etc.), and the equality in (\ref{eq:Palpha}) 
holds only when $\alpha$ is an even integer.
Moreover, there are rank-1 lattices point sets $P_n$ for which 
$\cP_\alpha$ converges as  $\cO(n^{-\alpha+\epsilon})$ for any $\epsilon > 0$
\cite{vDIC06a,vSIN12a,vSLO94a}.
Adding projection-dependent weights $\gamma_{\fraku}$ leads to the weighted 
${\cal P}_{\gamma,\alpha}$, defined by (\ref{eq:weightedFOM}) and (\ref{eq:FOMfraku}) 
with $q=2$,
\[
  \phi(u_{i,j}) = -(-4\pi^2)^{\alpha/2} B_{\alpha}(u_{i,j}) / \alpha!,
\]
and $\cD_{\fraku}^2(P_n) = {\cP}_{\alpha,\fraku}(P_n)$ is the $\cP_\alpha$
for the projection of $P_n$ on the coordinates in $\fraku$.

There is a similar variance expression for digital nets in base 2 with a random digital shift,
with the Fourier coefficients $\hat f({\bm h})$ replaced the the Walsh coefficients,
and the dual lattice replaced by the dual net \cite[Definition 4.76]{rDIC10a}
or the dual lattice in the case of PLRs \cite{vLEC04a,vLEM03a}.
Thus, FOMs that correspond to variance bounds can be obtained by finding easily computable 
bounds on the Walsh coefficients.
By assuming a rate of decrease of $\cO(2^{-\alpha |\bm h|})$
of the Walsh coefficients $\tilde f({\bm h})$ with 
${\bm h} = (h_1,\dots,h_s) \in \NN^s$ and $|{\bm h}| = |h_1| + \cdots + |h_s|$,
and using a RKHS with shift- and scramble-invariant kernel, 
\cite{vYUE02a} and \cite{vDIC05c} obtain a FOM of the form 
(\ref{eq:weightedFOM}) and (\ref{eq:FOMfraku}) with 
\begin{equation*}
 \phi(x) = \phi_{\alpha}(x) = \mu(\alpha) 
   - \II[x > 0] \cdot 2^{(1+\lfloor \log_2(x)\rfloor)(\alpha-1)}(\mu(\alpha) + 1),
\end{equation*}
where $\II$ is the indicator function,
$-\lfloor \log_2 x\rfloor$ is the index of the first nonzero digit in the expansion of $x$,
and $\mu(\alpha)= \left( 1-2^{1-\alpha} \right)^{-1}$ for any real number $\alpha>1$.
This gives $\mu(2) = 2$, $\mu(3) = 4/3$, \dots, 
For $\alpha=2$, this gives 
\[
  \phi_2(x) = 2 (1- \II[x > 0] \cdot 3 \cdot 2^{\lfloor \log_2(x)\rfloor}),
\]
which corresponds to the FOM suggested in \cite[Section 6.3]{vLEM03a} for PLRs.
In \cite{vHIC01d,vYUE02a}, $\phi(x)$ is written in terms of $\eta = \alpha-1$ instead,
but it is exactly equivalent.  These papers also show the existence of digital nets
for which the FOM converges as $\cO(n^{-\alpha} (\log n)^{s-1})$ for any $\alpha > 1$.
This FOM can be seen as a counterpart of $\cP_\alpha$ and we call it $\tilde{\cP}_\alpha$.
Its value $\tilde{\cP}_{\alpha,\fraku}(P_n)$ on the projection of $P_n$ on the 
coordinates in $\fraku$ can be used for $\cD_{\fraku}^2(P_n)$, with $q=2$.
Note that under our assumption that the first $k$ rows of each generating matrix
are linearly independent, $-\lfloor \log_2(u_{i,j}) \rfloor$ never exceeds $k$ when $u_{i,j} \not=0$,
and therefore this FOM depends only on the first $k$ bits of output.

Dick and Pillichshammer \cite[Chapter 12]{rDIC10a} consider a RKHS with shift-invariant
kernel, which is a weighted Sobolev space of functions whose mixed partial derivatives 
of order 1 are square-integrable.  
This gives a FOM of the form (\ref{eq:weightedFOM}) and (\ref{eq:FOMfraku}) with $q=2$ and 
\begin{equation*}
 \phi(x) = 1/6 -\II[x > 0] \cdot 2^{\lfloor \log_2(x)\rfloor -1}.
\end{equation*}
They show that there are digital nets for which this FOM (and therefore the square error)
converges almost as $\cO(n^{-2})$.
In their Chapter 13, they find that the scramble-invariant version gives the same $\phi$.
Note that this $\phi(x)$ is equal to $\phi_2(x)$ above, divided by 12.
Therefore, we can get the corresponding FOM just from $\tilde \cP_2$ by multiplying the weights
by order-dependent factors of $1/12^j$ for order $j$.

Goda \cite{rGOD15a} examines \emph{interlaced polynomial lattice rules} (IPLR),
also for a Sobolev space of order $\alpha$, with an interlacing factor $d > 1$.
He provides two upper bounds on the worst-case error in a deterministic setting.
These bounds can be used as FOMs.
The first is valid for all positive integer values of $\alpha$ and $d > 1$,
whereas the second holds only for $1 < d\le \alpha$, but is tighter when it applies.
These two bounds have the form (\ref{eq:weightedFOM}) and (\ref{eq:FOMfraku}) with $q=1$,
$\gamma_{\fraku}$ replaced by $\tilde\gamma_{\fraku}$, and
\[
 \phi(x_{i,j}) = -1 + \prod_{\ell=1}^d (1+\phi_{\alpha,d,\ell}(x_{i,(j-1)d+\ell})),
\]
where for the first bound, $\tilde\gamma_{\fraku} = \gamma_{\fraku} 2^{\alpha (2d-1)|\fraku|/2}$,
\[
	\phi_{\alpha,d,\ell}(x) 
	  = \frac{1 - 2^{(\min(\alpha,d)-1) \lfloor \log_2 x\rfloor} (2^{\min(\alpha,d)}-1)}
	               {2^{(\alpha+2)/2} (2^{\min(\alpha,d)-1}-1)}
\]
for all $x\in [0,1)$, where $2^{\lfloor \log_2 0\rfloor} = 0$,
while for the second bound, $\tilde\gamma_{\fraku} = \gamma_{\fraku}$ and
\[
	\phi_{\alpha,d,\ell}(x) = \frac{ 2^{d-1} (1 - 2^{(d-1) \lfloor \log_2 x\rfloor} (2^{d}-1))}
	               {2^{\ell} (2^{d-1}-1)}.
\]
One can achieve a convergence rate of almost $\cO(n^{-\min(\alpha,d)})$ for these FOMs
(and therefore for the worst-case error).
We denote these two FOMs by $\mathcal{I}_{\alpha,d}^{(a)}$ and $\mathcal{I}_{\alpha,d}^{(b)}$.

Goda and Dick \cite{rGOD15c} proposed another FOM, also for a Sobolev space of order $\alpha$,
for interlaced randomly-scrambled PLRs of high order.
They showed that this scheme can achieve the best possible convergence rate of 
$\cO(n^{-(2\min (\alpha,d) +1) +\delta})$ for the variance. 
The FOM, denoted ${\cal I}_{\alpha,d}^{(c)}$, has the same form, but with $q=2$,
\[
	\phi(x) = \phi_{\alpha,d}(x) 
	  = \frac{1 - 2^{2\min(\alpha,d) \lfloor \log_2 x\rfloor} (2^{2\min(\alpha,d)+1}-1)}
	               {2^\alpha (2^{2\min(\alpha,d)} -1)},
\]
and $\gamma_{\fraku}$ replaced by $\tilde\gamma_{\fraku} = \gamma_{\fraku}\, D_{\alpha,d}^{|\fraku|}$
where $D_{\alpha,d} = 2^{2\max(d-\alpha,0)+(2d-1)\alpha}$.

Another set of FOMs are obtained from upper bounds on the star discrepancy of $\cD^*(P_n)$ 
or its projections on subsets of coordinates, when $P_n$ is a digital $(t,k,s)$-net.
One such bound is $\cD^*(P_n) \le 1 - (1-1/n)^s + \cR_2$ where
\begin{equation}
\label{eq:starDboundR2}
  \cR_2 = -1 + \frac{1}{n} \sum_{i=0}^{n-1} \prod_{j=1}^s 
		       \left[\sum_{k=0}^{n-1} 2^{-\lfloor \log_2 k\rfloor -1} {\rm wal}_k(u_{i,j})\right]				
\end{equation}
${\rm wal}_k$ is the $k$th Walsh function in one dimension, and
we assume that the generating matrices ${\bm C}_j$ are $k\times k$.
See \cite[Theorems 5.34 and 5.36]{rDIC10a}, where a more general version with projection-dependent 
weights is also given.   
For PLRs in base $b=2$, this criterion is equal to $\cR'_{2,\gamma}$ given in \cite[Chapter 10]{rDIC10a}:
\begin{equation}
\label{eq:starDboundR2PLR}
  \cR'_{2,\gamma} = -\sum_{\emptyset\not=\fraku\subseteq\{1,2,\dots,s\}} \gamma_{\fraku}
	           + \frac{1}{n} \sum_{i=0}^{n-1} \sum_{\emptyset\not=\fraku\subseteq\{1,2,\dots,s\}}
	             \gamma_{\fraku} \prod_{j\in\fraku} \phi_k(u_{i,j})
\end{equation}
where $\phi_k(u) = -\lfloor \log_2(u) \rfloor/2$ if $u \ge 2^{-k}$ and $\phi_k(u) = 1+ k/2$ otherwise.

A classical upper bound on the star discrepancy is also given by the $t$-value of the digital net:
\[
  \cD^*(P_n) \le \frac{2^t}{n} \sum_{j=0}^{s-1} \binom{k-t}{j}.
\]
If we use this upper bound for each projection on the subset $\fraku$ of coordinates, 
we get the FOM (\ref{eq:weightedFOM}) with $q=1$ and
\[
  \cD_{\fraku}(P_n) = \frac{2^{t_{\fraku}}}{n} \sum_{j=0}^{|\fraku|-1} \binom{k-t_{\fraku}}{j}
\]
where $t_{\fraku} = t_{\fraku}(P_n)$ is the $t$-value of the projection of $P_n$ on the 
coordinates in $\fraku$.
LatNet Builder implements this with arbitrary weights, 
using algorithms described in \cite{rMAR20a}.
Dick \cite{vDIC08a} obtains worst-case error bounds that converge at rate almost 
$\cO(n^{-\alpha})$ for interlaced digital nets, based on the $t$-values of the projections.

\section{Search Methods} 
\label{sec:search-methods}

For given construction type, FOM, and weights, finding the best choice of parameters 
may require to try all possibilities, but their number is usually much too large.
LatNet Builder implements the following search methods.

In an \emph{exhaustive search}, all choices of parameters are tried, so we are guaranteed
to find the best one.  
This is possible only when there are not too many possibilities.

A \emph{random search} samples uniformly and independently a fixed number $r$ of parameter choices,
and the best one is retained.

In a \emph{full component-by-component (CBC)} construction, the parameters are selected for one 
coordinate at a time, by taking into account the choices for the previous coordinates only
\cite{vDIC05b,rDIC10a,vSLO02c}.
The parameters for coordinate $j$ (e.g., the $j$th coordinate of the generating vector in the 
case of lattices), are selected by minimizing the FOM for the first $j$ coordinates, in $j$ dimensions,
by examining all possibilities of parameters for this coordinate, without changing the parameter choices
for the previous coordinates.  This is done for the $s$ coordinates in succession.
This greedy approach can reduce by a huge factor (exponential in the dimension) the total number 
of cases that are examined in comparison with the exhaustive search. 
What is very interesting is that for most types of QMC constructions and FOMs
implemented in LatNet Builder, 
the convergence rate for the worst-case error or variance obtained with this restricted approach 
is provably the same as for the exhaustive search \cite{rDIC10a,rGOD19a}.

For lattice-type point sets, with certain FOMs and choices of weights
(e.g., $\cP_2$ and $\tilde \cP_2$ with product and/or order-dependent weights), 
a \emph{fast CBC} construction 
can be implemented by using a fast Fourier transform (FFT), so the full CBC construction
can be performed much faster \cite{rDIC10a,rNUY06b,rNUY14a}.
LatNet Builder supports this.

When the number of choices for each coordinate is too large or fast-CBC does not apply,
one can examine only a fixed number of random choices for each coordinate $j$;
we call this the \emph{random CBC} construction.

For lattice-type constructions, one can also further restrict the search to 
\emph{Korobov}-type generating vectors.
The first coordinate is set to 1 and only the second coordinate needs to be selected.
This can be done either by an exhaustive search or by just taking a random sample for the 
second coordinate (\emph{random Korobov}).

For digital nets, a \emph{mixed CBC} method is also available: it uses full CBC for the first
$d-1$ coordinates and random CBC for the other ones, for given $1\le d \le s$.

\section{Usage of the Software Tool}
\label{sec:software}

At the first level, LatNet Builder is a library written in C++ which implements classes and methods to compute
FOMs and search for good point sets for all the construction methods and FOMs discussed in this paper.
The source code and a detailed reference manual for the library are provided at
\url{https://github.com/umontreal-simul/latnetbuilder}.
The library can be used directly from C++ programs and can be extended if desired.
This is the most flexible option, but it requires knowledge of C++ and the library.

At the second level, there is an executable \texttt{latnetbuilder} program that can be called 
directly from the command line in Linux, Mac OS, or Windows.
This program has a large number of options to specify the type and number of points, number of dimensions, 
search method, FOM, weights, output file format, etc. 
We think this is the most convenient way of using LatNet Builder in practice.
In addition to giving the search results on the terminal, the program creates a directory with two output files: 
one summarizes the search parameters and the other one puts the parameters of the selected point set 
in a standard format designed for reading by
other software that can generate and use the RQMC points in applications.
There are selected file formats for ordinary lattice rules, polynomial lattice rules,
Sobol points, and general digital nets in base 2.
A tutorial on the command line and a summary of the options can be found in the reference manual.

At a third level, there is a Java interface in SSJ, a Python interface included in the distribution, and 
a Graphical User Interface (GUI) based on the Jupyter ecosystem, written in Python.
These interfaces use the command line internally. 
With the GUI, the user can select the desired options in menus, write numerical values in input cells, 
write the name of the desired output directory, and launch a search.
The LatNet Builder program with the Python interface and the GUI can be installed as a Docker container on one's machine.
An even simpler access to the GUI is available without installing anything: just click on the 
``Launch Binder'' black and pink link on the GitHub site and it will run the GUI with a version of the program 
hosted by Binder. This service provides limited computation resources but is convenient for small experiments
and to get a sense of what the software does.

We now give examples to illustrate how the command line works, how the results look like,
and give some idea of the required CPU times for the search.
The timings were made in a VirtualBox for Ubuntu Linux running atop Windows 10 on an old desktop 
computer with an Intel i7-2600 processor at 3.4GHz and 32 Gb of memory.

The following command makes a search for a polynomial lattice rule with $n=2^{16}$ points in 256 dimensions, 
with the default irreducible polynomial modulus,
using the fast-CBC search method, 
the $\tilde P_2$ criterion, $q=2$, and order-dependent weights
$\Gamma_2 = 10$, $\Gamma_3 = 0.1$, $\Gamma_4 = 0.001$, and the other weights equal to 0
(recall that $\Gamma_1$ has no impact on the selection). 
These weights decrease quickly with the order because the number of projections of any given order
increases very quickly with the order.  If they decrease too slowly, the total weight of the 
projections of order 2 in the FOM will be negligible compared to those of order 4, for instance.
For a smaller $s$, the weights may decrease more slowly.

\smallskip\smallcode
latnetbuilder -t lattice -c polynomial -s 2^16 -d 256 -e fast-CBC 
    -f CU:P2 -q 2 -w order-dependent:0,0,10.,0.1,0.001 -O lattice
\endcode
\smallskip

This search takes about 840 seconds to complete and the retained rule had an FOM of 60.235.
About 70\% of this FOM is contributed by the projections of order 4 and less than 1\% by the 
projections of order 2.  The output file looks like:

\smallskip
\smallcode
   # Parameters for a polynomial lattice rule in base 2
   256     # s = 256 dimensions
   16      # n = 2^16 = 65536 points
   96129   # polynomial modulus
   1       # coordinates of generating vector, starting at j=1
   47856
   60210
   44979
     \vdots
\endcode

Instead of making the search directly in the polynomial lattice space, 
we can make the same search by viewing the PLR as a digital net, using the option ``{\tt -t net}.''  
With that option, fast-CBC search is not available, 
but we can do a random CBC search, say with 100 samples for each coordinate.
With either search method, instead of reporting the modulus and generating vector of the PLR 
in the output file, we can have all the columns of the generating matrices, 
by using the option ``{\tt -O net}'' instead of ``{\tt -O lattice}.''  The command is:

\smallskip
\smallcode
latnetbuilder -t net -c polynomial -s 2^16 -d 256 
        -e random-CBC:100 -f CU:P2 -q 2 
        -w order-dependent:0,0,10.,0.1,0.001 -O net
\endcode
\smallskip

\noindent
With this, the search takes about 160 seconds and gives an FOM of 63.25.
The output file looks like this, with 16 integers per dimension,
one integer for each column:

\smallskip
\smallcode
   # Parameters for a digital net in base 2
   256  # s = 256 dimensions
   16   # n = 2^16 = 65536 points
   31   # r = 31 binary output digits
   # Columns of gen. matrices C_1,...,C_s, one matrix per line:
   33260 66520 133040 266081 532162 1064325 2128651 4257303  ... 
   1561357389 975231131 1950462262 1753440876 1359398105 ...
   1642040599 1136597551 125711455 251422911 502845823 ...
     \vdots
\endcode
\smallskip

For a search in 32 dimensions instead of 256, it takes about 40 seconds for the first case 
and 8 seconds for the second case.

As another example, the following command launches a search for good direction numbers for 
Sobol' points for up to $2^{16}$ points in 256 dimensions.
It uses a mixed CBC search which does a full CBC evaluation for the first 10 coordinates and
then a random-CBC search with 100 random samples for each of the remaining coordinates.
The criterion is the maximum $t$-value with order-dependent weights of $\Gamma_2 = 1.0$, $\Gamma_1 = 0.5$,
and 0 for everything else.  Here, since we take the sup over the projections, the weights can decrease
much more slowly.  

\smallskip
\smallcode
latnetbuilder -t net -c sobol -s 2^16 -d 256 -e mixed-CBC:100:10  
     -f projdep:t-value -q inf -w order-dependent:0:0,1.0,0.5
\endcode
\smallskip

\noindent
The search took about 3340 seconds and returned a FOM of 8.0.
The output file provides the retained direction numbers 
and it looks like this:

\smallskip
\smallcode
   # Initial direction numbers m_{j,c} for Sobol points
   # s = 256 dimensions
   1    # This is m_{j,k} for the second coordinate 
   1 1
   1 1 1
   1 1 1
   1 1 5 1
   1 3 5 1
   1 1 1 9 9
   1 1 1 9 17
      \vdots
\endcode	
\smallskip

If we change $s$ to 32, the search takes 12 seconds and the FOM is 5.0 instead.

\section{Simple Numerical Illustrations}
\label{sec:illustrations}

Here we give a few simple examples of what LatNet Builder can do.
The simulation experiments, including the generation and randomization of the points, 
were done using SSJ \cite{sLEC05a}.

\subsection{FOM quantiles for different constructions}

\begin{figure}
\caption{The 0.1, 0.5, and 0.9 quantiles of the FOM distribution as functions of $n$ 
  for various constructions, in log-log scale.}  
\label{figure:convergence-FOM-distribution}
\centering
         \includegraphics[width=0.8\textwidth]{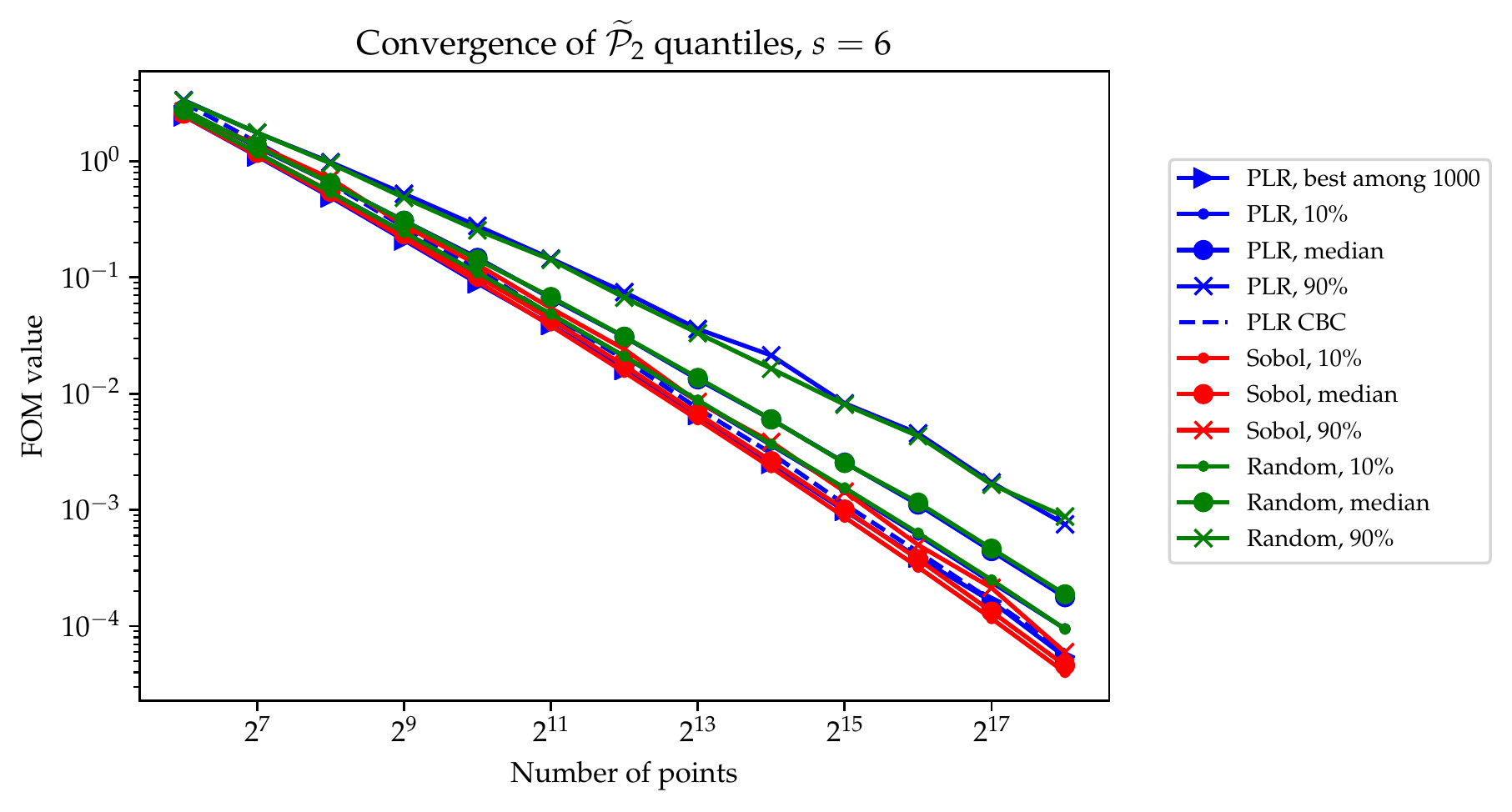}
         \includegraphics[width=0.8\textwidth]{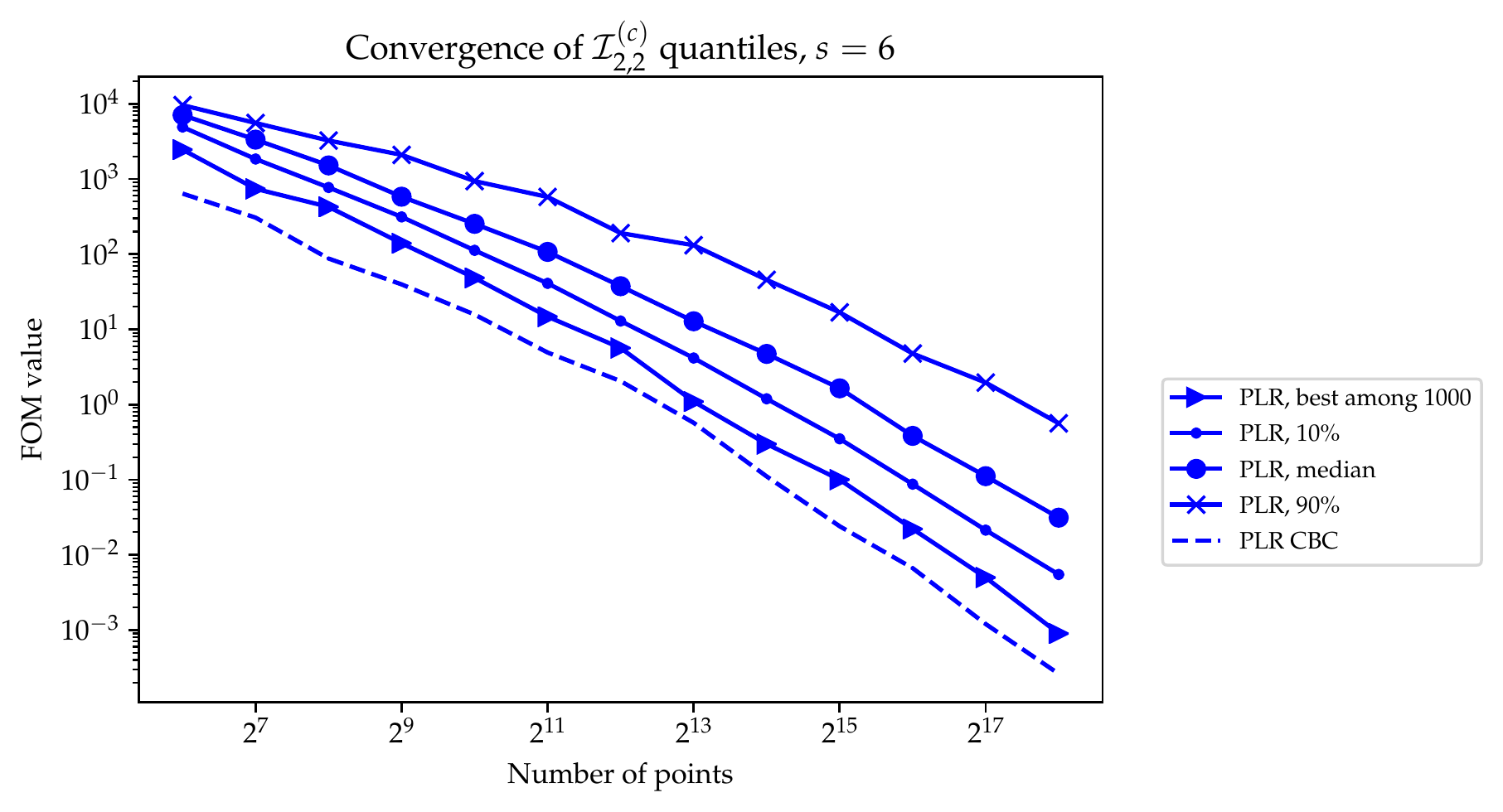}
\end{figure}

One might be interested in estimating the probability distribution of FOM values
obtained when selecting parameters at random for a given type of construction,
perhaps under some constraints, and as a function of $n$.
Here we estimate this distribution by its empirical counterpart with an independent sample of size 1000 (with replacement), 
and we report the 0.1, 0.5 and 0.9 quantiles of this empirical distribution, 
for $n$ going from $2^6$ to $2^{18}$.  
We do this for PLRs, Sobol' points, and digital nets with arbitrary invertible 
and projection-regular generating matrices (random nets), with $\tilde{\mathcal{P}}_2$ taken 
as the FOM, in $s=6$ dimensions, with $\gamma_{\fraku} = 0.7^{|\fraku|}$ for all $\fraku$.
We also report the value obtained by a (full) fast CBC search for a PLR.
The results are displayed in the first panel of Figure~\ref{figure:convergence-FOM-distribution}.
We see that the FOM distribution has a smaller mean and much less variance for the Sobol' points than for 
the other contructions. 
Even the median obtained for Sobol' beats (slightly) the FOM obtained 
by a full CBC construction with PLRs.
The quantiles for random PLRs and random nets are approximately the same.

The second panel of the figure shows the results of a similar sampling for PLRs with 
${\cal I}_{2,2}^{(c)}$ 
as a FOM, also in 6 dimensions.  
Here, the FOM values are more dispersed and the fast CBC gives a significantly better value than the best FOM obtained by random sampling. Also the search for the point set parameters is much quicker with the fast CBC construction than with random sampling of size 1000.

\subsection{Comparison with tabulated parameter selections}

\begin{figure}
\caption{Distributions of $t$-values for 2-dim and 3-dim projections, for three Sobol' point sets: 
(1) \emph{Joe-Kuo} taken from \cite{rJOE08a}, 
(2) \emph{Max} and (3) \emph{Sum} are found by LatNet Builder as explained in the text.
For each case, we report the number of projections having any given $t$-value, 
as well as the average $t$-value (dashed vertical lines).}
\label{figure:comparison-joe-kuo}
\centering
     \begin{subfigure}[b]{0.49\textwidth}
         \centering
         \includegraphics[width=\textwidth]{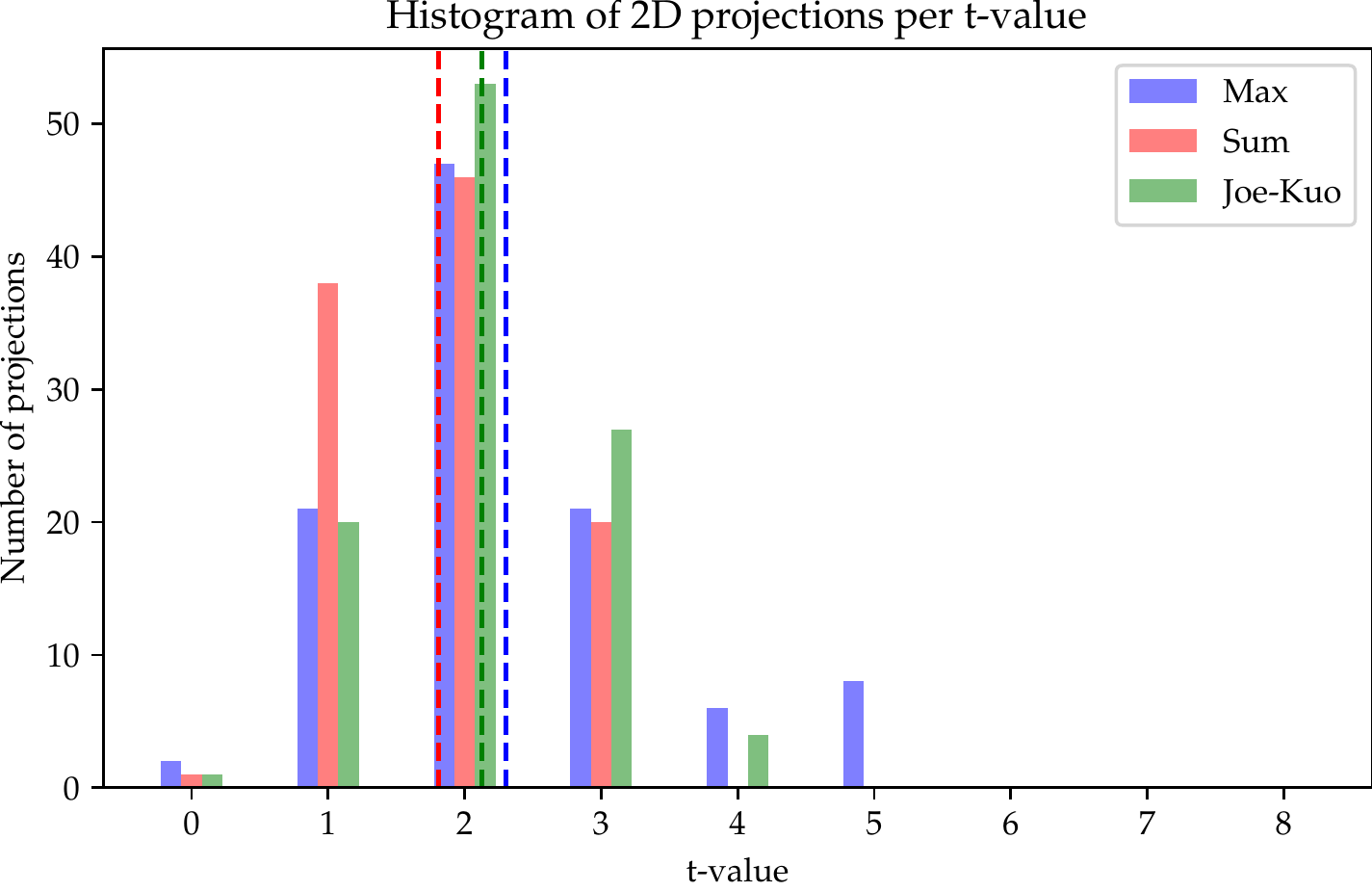}
     \end{subfigure}
     \hfill
     \begin{subfigure}[b]{0.49\textwidth}
         \centering
         \includegraphics[width=\textwidth]{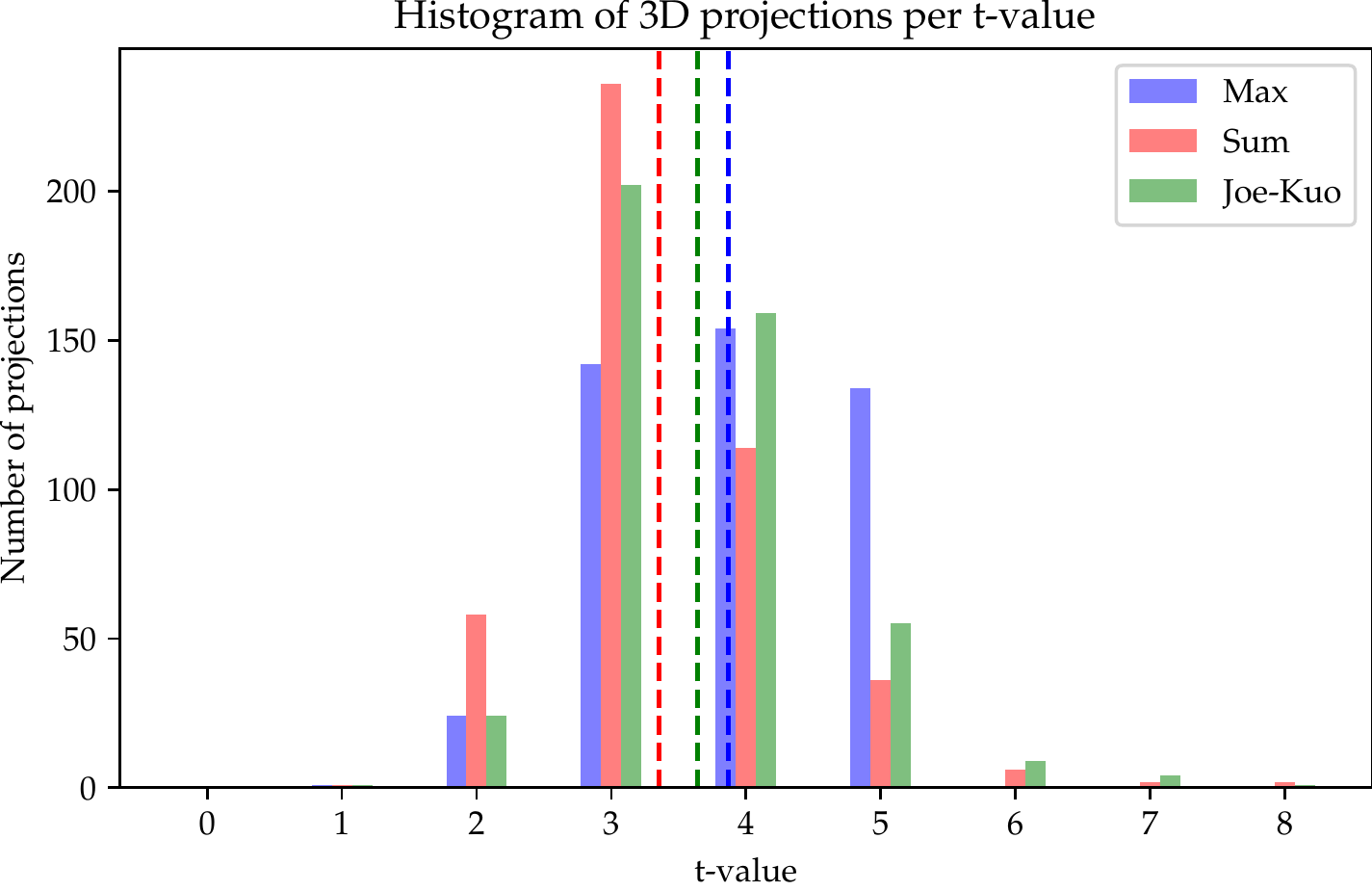}
     \end{subfigure}
\end{figure}

We now give small examples showing how searching for custom parameter values with LatNet Builder 
can make a difference in the RQMC variance compared with pre-tabulated parameter values 
available in software or over the Internet.
We do this for Sobol' nets, and our comparison is with the precomputed direction numbers
obtained by Joe and Kuo \cite{rJOE08a}, which are arguably the best proposed values so far.
These parameters were obtained by optimizing a FOM based on the $t$-values over two-dimensional 
projections, using a CBC construction.
With LatNet Builder, we can account for any selected projections in our FOM.
For instance, if we think all the projections in two and three dimensions are important,
we can select a FOM that accounts for all these projections.
To illustrate this, we made a CBC construction of $n = 2^{12}$ Sobol' points in $s=15$ dimensions,
using the sum or the maximum of $t$-values in the two- and three-dimensional projections.
Figure \ref{figure:comparison-joe-kuo} shows the distribution of $t$-values obtained
with the sum, the max, and the points from \cite{rJOE08a}.
Compared with the latter, we are able to reduce the worse $t$-value over 3-dim projections 
from 8 to 5 when using the max, and to reduce the average $t$-value when using the sum.
However, when using the max, we get a few poor two-dim projections, because we compare the 
$t$-values on the same scale for two and three dimensions. 
We should probably multiply the $t$-value by a scaling factor 
that decreases with the dimension.

In our next illustration, we compare the RQMC variances for Sobol' points with direction
numbers taken from \cite{rJOE08a} and direction numbers found by LatNet Builder using a custom
FOM for our function. We want to integrate
\begin{equation*}
 f(\bm u) = \prod_{j=1}^{5}(\psi(u_j) - \mu) + \prod_{j=6}^{10}(\psi(u_j) - \mu),
\end{equation*}
where $\psi(u) = \left((u - 0.5)^2 +0.05 \right)^{-1}$
and $\mu=\EE[\psi(U)]\approx10.3$ when $U\sim U(0,1)$.
This function is the sum of two five-dimensional ANOVA terms for a more general function 
taken from \cite{iGEN84a}.
A good FOM for this function should focus mainly on these two five-dim projections,
namely $\mathfrak{u} = \{1,2,3,4,5\}$ and $\mathfrak{u} = \{6,7,8,9,10\}$,
and not on the two-dim projections as in \cite{rJOE08a}.
So we made a search with the $\tilde{\cP}_2$ criterion
with weights $\gamma_{\mathfrak{u}} = 1$ for these two projections and 0 elsewhere,
to obtain new direction numbers for $n = 2^{20}$ Sobol' points in 10 dimensions.
Then we estimated the variance of the sample RQMC average over these $n$ points
with the two choices of direction numbers 
(those of \cite{rJOE08a} and ours), using $m=200$ independent replications of an RQMC scheme
that used only a random digital shift.
The empirical variance with our custom points was smaller by a factor of more than 18.

\subsection{Variance for another toy function}

Here we consider a family of test functions similar to those in \cite{rSOB03a}:
\[
	f_{s, {\bm c}}({\bm u}) = \prod_{j=1}^s (1 + c_j \cdot (u_j-1/2))
\]
for ${\bm u} \in (0, 1)^s$, where ${\bm c} = (c_1,\dots,c_s) \in (0, 1)^s$.
The ANOVA components are, for all $\mathfrak{u} \subset\{1,\dots,s\}$,
\[
  (f_{s, {\bm c}})_\mathfrak{u}({\bm u}) = \prod_{j \in \mathfrak{u}} c_j \cdot (u_j-1/2)).
\]

\begin{figure}
\caption{Variance as a function of $n$ in log-log scale, for PLRs with two randomization schemes 
  and three interlacing factors $d$, found with LatNet Builder.  We also report the average time to 
	generate and randomize the points with LMS+shift.}
\label{figure:variance-conv}
\centering
\begin{minipage}[c]{0.70\linewidth}
\hskip-12pt
\includegraphics[width=\textwidth]{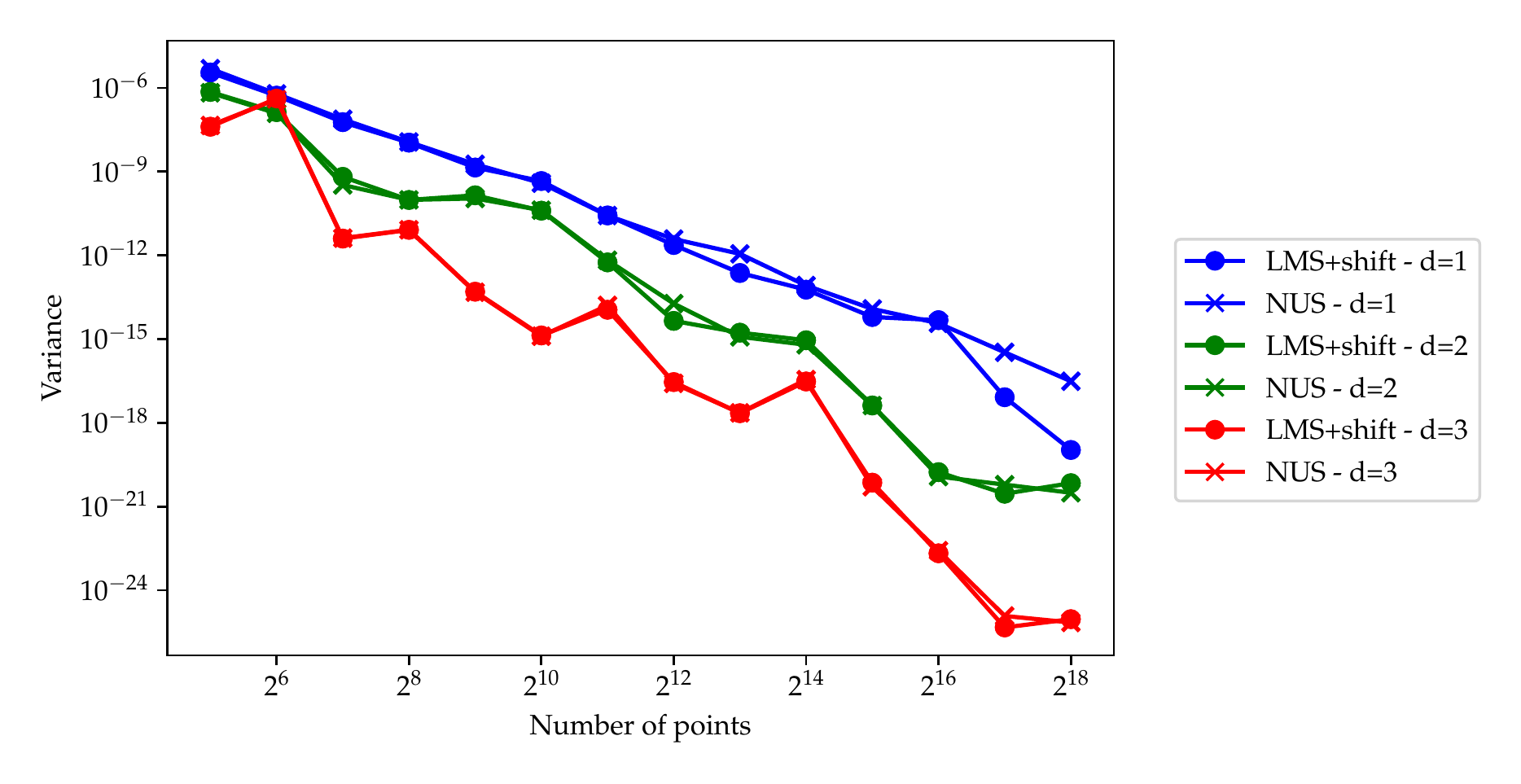}
\end{minipage}
\begin{minipage}[c]{0.28\linewidth}
\begin{tabular}{|c|c|}
\toprule
Nb. of points &     Constr. time \\
\midrule
128       &            0.02 s  \\
512       &            0.04 s \\
2048      &            0.12 s \\
8192      &            0.50 s  \\
32768     &            2.25 s  \\
131072    &            8.32 s \\
\bottomrule
\end{tabular}
\end{minipage}
\end{figure}

For an experiment, we take arbitrarly $s=3$ and ${\bm c} = (0.7, 0.2, 0.5)$.
We use LatNet Builder to search for good PLRs with a fast CBC construction, 
with product weights $\gamma_j = c_j$, with the FOMs 
$\mathcal{P}_2$, $\mathcal{I}^{(c)}_{2, 2}$, and $\mathcal{I}^{(c)}_{3, 3}$ 
(whose interlacing factors $d$ are 1, 2, and 3, respectively). 
For each $n=2^k$, $k = 5, \dots, 18$, we estimate the RQMC variance with $m$ independent 
replications of the randomization scheme, with $m = 1000$ for LMS+shift, and $m=100$ for NUS.  
For the interlaced points, the randomization is performed before the interlacing, as in \cite{rGOD15c}. 
Figure~\ref{figure:variance-conv} shows the variance as a function of $n$, in log-log scale.
We see that the two randomization schemes give approximately the same variance.
However, the time to generate and randomize the points is much larger for NUS than for LMS+shift: 
around 10 times longer for $2^{11}$ points and 50 times longer for $2^{18}$ points. 
As expected, the variance reduction and the convergence rate are larger when the interlacing factor 
increases, although the curves are more noisy. 

\section{Conclusion}
\label{sec:conclusion}

LatNet Builder is both a tool for researchers to study the properties of highly uniform point sets and associated figures of merit, and for practitioners who want to find good parameters for a specific task.
It is relatively easy to incorporate new FOMs into the software, 
especially if they are in the kernel form (\ref{eq:FOMfraku}).

Many questions remain open regarding the roles of the construction, the search method, 
the randomization, and (perhaps more importantly) the choice of the weights. 
It is our hope that the software presented here will spur interest into these issues.

\begin{acknowledgement}
This work has been supported by a NSERC Discovery Grant
and an IVADO Grant to P. L'Ecuyer, and by a stipend from Corps des Mines to P. Marion.
F. Puchhammer was supported by Spanish and Basque governments fundings through BCAM (ERDF, ESF, SEV-2017-0718, PID2019-108111RB-I00, PID2019-104927GB-C22, BERC 2018e2021, EXP. 2019/00432, KK-2020/00049), and the computing infrastructure of i2BASQUE  and IZO-SGI SGIker (UPV).
Yocheved Darmon wrote code for producing the output files.
\end{acknowledgement}
\vskip-25pt

%
\bibliographystyle{spmpsci}
\bibliography{vrt,random,math,stat,simul,ift,optim,fin}

\end{document}